# Beyond quantum microcanonical statistics

**Barbara Fresch[1], Giorgio J. Moro[1]**

*Dipartimento di Science Chimiche, Università di Padova,*

*via Marzolo 1, 35131 Padova, Italy*

## Abstract

*The Schrödinger equation for the evolution of isolated quantum systems determines the constants of motion for the dynamical problem, which can be identified with the populations along the principal directions of the Hamiltonian. The microcanonical statistics assumes from the beginning fixed values for the populations, while a more realistic view of material systems should prescribe a distribution with respect to all the possible populations. Such a more general statistical description of quantum systems is introduced in the present work within the Random Pure State Ensemble (RPSE) as obtained from a random homogeneous distribution of the wavefunctions on the unit sphere of the active Hilbert space. From the corresponding probability distribution on populations the typicality is verified for microscopic (i.e., dependent on the populations) equilibrium properties like the time average of the reduced density matrix of a subsystem, the expectation value of the hamiltonian and the Shannon entropy with respect to the populations. This allows the identification, through the RPSE average, of macroscopic properties which are independent of the specific realization of the quantum system. A description of material systems in agreement with equilibrium thermodynamics is then derived without constraints on the physical constituents and interactions of the system. Furthermore, the canonical statistics is recovered for the typical equilibrium reduced density matrix of a subsystem.*

---

[1] E-mails: barbara.fresch@unipd.it; giorgio.moro@unipd.it

# I. Introduction

The standard treatment of the statistical thermodynamics (mechanics) of isolated quantum system relies on the introduction of the microcanonical density matrix. It has a very peculiar structure: it is vanishing outside the subspace spanned by the Hamiltonian principal directions with eigenvalues within $E_{\min}$ and $E_{\max}$ for a sufficiently small (but not too much) energy width $\Delta E = E_{\max} - E_{\min}$, and its projection onto such a subspace is proportional to the unity matrix. On the other hand a quantum dynamic process is fully characterized by solving the time-dependent Schrödinger equation for a given initial wavefunction (pure state) [1,2].

As a matter of fact, these two levels of descriptions are not satisfactorily related on a logical ground. In particular, the introduction of the microcanonical density matrix are not a direct and necessary result of the unitary (Schrödinger) evolution of pure states, rather it is often postulated on the basis of an analogy with the classical microcanonical distribution [3].

The aim of the present paper is the complete characterization of equilibrium properties of a generic isolated system from a completely different standpoint: rather then assuming the microcanonical statistical density matrix as describing the equilibrium state of an isolated quantum system we start from the idea that an isolated quantum system is described by its time evolving wavefunction. The connection with thermodynamics is then built by means of a statistical analysis of the possible pure states of the system. The idea of assigning probability distribution to wavefunctions has been used in the past to calculate on a statistical basis molecular properties such as transition probabilities [4,5,6] and to characterized vibrationally excited states of polyatomic molecules [7,8,9]. This was suggested by the observation that quantum states describing molecular excited states are combination of many eigenstates of a "zeroth-order" (without interaction terms) molecular Hamiltonian [10] and thus the formulation of a probability distribution on the coefficients of such an expansion permits simple evaluation of average molecular properties. More recently, the general procedure of assigning probabilities to pure states on the basis of the volume in the corresponding Hilbert Space has been employed to clarify foundational aspects of quantum statistical mechanics [11,12] as well as to study relaxation from a general perspective [13]. An important step in the study of quantum states from a statistical standpoint has been the recognition that many quantum properties manifest typicality [14,15]. In its widest meaning the term typicality indicates that by selecting a set of pure states on the basis of some conventional statistical rules, one obtains a very narrow distribution of some relevant features which, therefore, become typical amongst the possible pure states. Following this line of reasoning it has been shown in Ref. [11] that a typical reduced density matrix for a subsystem arises from the overwhelming majority of pure states belonging to the Hilbert space or to a Hilbert subspace selected by some constraints. Independently Goldstein et al. demonstrated [12] that if one considers the set of the pure states which are superpositions of the Hamiltonian eigenstates corresponding to the energy shell $\Delta E$ usually associated with the microcanonical statistics, then the corresponding typical reduced density matrix is of the standard canonical form. These results have introduced a new paradigm in quantum statistical mechanics: from the traditional point of view assuming a uniquely defined statistical density matrix, to the probabilistic analysis of a single quantum system described by its wavefunction. In subsequent contributions [14] it has been recognized that typicality characterizes a general class of observables and not only the state of a subsystem. Moreover there are now many attempts to push the same approach beyond the description of the equilibrium state in order to gain insight into the dynamical problem [13,16,17].

We have developed [18] a description of the equilibrium of quantum systems which, on the one hand, shares the importance of statistical typicality with the above mentioned contributions but, on the other hand, it attributes a privileged role to the quantum dynamics in determining the statistical description of the isolated systems. The starting point of our theory is the statistical characterization of the time dependent wavefunction which described an isolated quantum system. Once a suitable parameterization has been introduced for the wavefunction, from the analysis of its temporal evolution one can derive the distribution on these parameters, which is called Pure State Distribution (PSD). In such an operation it is convenient to employ as parameters the set of phases and populations deriving from the polar representation of each component of the wavefunction along the principal directions of the Hamiltonian. Indeed, one can show under rather mild conditions [18], that the populations are constant of motion fixed by the initial condition, while the phases are uniformly distributed in their standard domain. In order to characterized the equilibrium we consider two categories of functions of the quantum state: i) collective properties of the isolated system, like the internal energy calculated as expectation value of the Hamiltonian and the (Shannon) entropy determined by the populations, and ii) properties of a subsystem which can be evaluated according to the corresponding reduced density matrix.

In such a framework, the standard microcanonical statistics can be recovered as the time average of the instantaneous pure state density matrix, only once a very particular set of populations has been selected [18], However, quantum dynamics does not provide any information on the populations, as long as they are constants of motion. Therefore arbitrary values can be attributed to these parameters, once the population's normalization is taken into account, together with further constraints [19,20,21] dictated, for instance, by the need of selecting the thermodynamical internal energy of the isolated system. The same quantum system can have different realizations characterized by different sets of populations, and there are no reasons to privilege a priori a particular set of populations amongst the admissible ones.

Because of the lack of information about the populations, one can characterize them only in a statistical sense. More precisely, just because no privilege can be attributed to any particular set of populations, one assigns an identical statistical weight to the quantum states belonging to the unitary hyper-sphere of the Hilbert space representative of normalized wavefunctions, or to its lower dimensional subspaces deriving from the imposition of further constraints. The resulting probability density on the populations, together with the sample space taking into account the constraints, define the statistical ensemble [22] for the population set.

However, different ensembles can be proposed in relation to the constraints to be enforced, and a priori there are not obvious criteria of choice. In [18] we proposed that some fundamental requirements have to be satisfied by the population statistics. The internal energy and the entropy must display typicality in the large size limit of systems at constant volume, leading to a macroscopic equation of state for the internal energy dependence $S(U)$ of the entropy in agreement with the thermodynamical behaviour of material systems. In the same limit typicality has to be recovered for the reduced density matrix in correspondence of the canonical form at the temperature defined by the entropy equation of state.

In ref [18] we have introduced a particular ensemble, called Random Pure State Ensemble (RPSE), with population statistics in agreement with a random choice of the wavefunction belonging to an active Hilbert subspace defined on the basis of an upper cut-off energy. Such a cut-off energy is necessary in order to deal with finite dimensional statistics. We have shown that RPSE supplies an appropriate ensemble for the populations in the particular case of an ideal system of (non-interacting) identical spins with $J=1$ quantum number, in the meaning that it leads to a coherent thermodynamical characterization of such an isolated spin system. We mention that in the same work a different ensemble, the Fixed Expectation Energy Ensemble based on the constraint of fixed expectation energy, was also analyzed but deriving that it does not lead to the emergence of well-behaving thermodynamic functions. Such a result illustrates how the above mentioned requirements based on the agreement with standard thermodynamics discriminate different ensembles.

In our opinion such an approach provides a description of material systems more profound and effective than the conventional microcanonical ensemble. However, in Ref. [19] it has been validated only for the very particular system of non interacting spins. The purpose of the present work is that of demonstrating its generality by verifying that it can applies to all the material systems without constraints on their physical constituents or interactions.

The rest of the paper is organized as follow. In the next section the Pure State Distribution describing isolated quantum systems is introduced together with the definition of the equilibrium properties and the microscopic definitions of internal energy and entropy. In section III the statistics of populations is characterized according to the Random Pure State Ensemble. This allows us to demonstrate in all generality the typicality of the microscopic entropy and internal energy, on the one hand and of the equilibrium reduced density matrix of a subsystem, on the other hand, under the only hypothesis of a large dimension of the active Hilbert space. These results for the typicality allow in the next section the identification of macroscopic entropy and internal energy behaving in agreement with thermodynamics. Furthermore, the canonical form is recovered for the typical reduced density matrix. The final section reports some general remarks about our methodological choices.

**II. Pure State Distribution (PSD)**

We consider a generic isolated quantum system characterized by its wavefunction $|\psi(t)\rangle$ belonging to the Hilbert space $\mathcal{H}$, whose time dependence is ruled by the Schrödinger equation for the given Hamiltonian $H$

$$|\psi(t)\rangle = \exp(-iHt/\hbar)|\psi(0)\rangle \tag{1}$$

with normalization $\langle\psi(t)|\psi(t)\rangle = 1$. In order to deal with its explicit time dependence, in the following we shall employ the wavefunction decomposed along the principal directions $|E_k\rangle$ of the Hamiltonian

$$H|E_k\rangle = E_k|E_k\rangle \tag{2}$$

for $k = 1, 2, \cdots$, with $\langle E_k|E_{k'}\rangle = \delta_{k,k'}$. We assume that the energy eigenvalues $E_k$ are rationally independent [23]. In real systems, with different types of interactions, each of them with a different magnitude according to the interparticle distance, the energy eigenvalues are characterized by a distribution with at least a partially random character [24,25,26,27]. This is the underlying point of view which supports the statistical analysis of the energy levels in complex quantum system [24,28,29] and the employment of mathematical tools like the random matrix [24] to model generic interactions. As a consequence the rational independence of the eigenenergies is a quite natural and weak restriction for real systems. In particular this implies that the energy eigenvalues are not degenerate and, therefore, they can be ordered in magnitude as $E_k < E_{k+1}$. Furthermore, in order to deal with a finite set parameterization of the wavefunction, we assume that $|\psi(t)\rangle$ belongs to the finite dimensional subspace $\mathcal{H}_N \subseteq \mathcal{H}$ in the following called as the active Hilbert space (for the wavefunction), and defined on the basis of the cutoff energy $E_{\max}$

$$\mathcal{H}_N := \text{span}\{|E_k\rangle | E_k < E_{\max}\} \tag{3}$$

where $N$ is its dimension: $E_N < E_{\max} \leq E_{N+1}$. The role of the cutoff energy, besides being a necessary ingredient in order to deal with a finite dimensional statistics, will appear clear when the thermodynamic limit is considered.

Under these hypotheses, the time dependence of the wavefunction is naturally parameterized according to its projections along the Hamiltonian principal directions

$$|\psi(t)\rangle = \sum_{k=1}^{N} c_k(t) |E_k\rangle \tag{4}$$

with coefficients

$$c_k(t) := \langle E_k | \psi(t) \rangle = \exp(-iE_k t / \hbar) c_k(0) \tag{5}$$

Let us introduce the following polar representation of the time dependent coefficients

$$c_k(t) = \sqrt{P_k} \exp\{-i\alpha_k(t)\} \tag{6}$$

with phases $\alpha_k$ linearly dependent on the time

$$\alpha_k(t) = \alpha_k(0) + E_k t / \hbar \tag{7}$$

and constant squared norms $P_k$, in the following denoted as populations,

$$P_k := |c_k(t)|^2 = |c_k(0)|^2 \tag{8}$$

normalized as

$$\sum_{k=1}^{N} P_k = 1 \tag{9}$$

The wavefunction can then be parameterized according to the set of time dependent phases, $\alpha = (\alpha_1, \alpha_2, \cdots, \alpha_N)$, and the set of populations, $P = (P_1, P_2, \cdots, P_N)$, which represents the constants of motion for the Schrödinger dynamics. Correspondingly any property of the quantum state can be represented as a function of the phases $f_P(\alpha(t))$ parametrically dependent on the constants of motion $P$, with a well defined asymptotic time average

$$\overline{f_P} := \lim_{T \to \infty} \frac{1}{T} \int_0^T dt \, f_P(\alpha(t)) \tag{10}$$

On the other hand, the phases can be considered as stochastic variables whose time dependence leads to a probability distribution (probability density) $p(\alpha)$ which allows the calculation of the time average

$$\overline{f_P} = \int d\alpha\, f_P(\alpha) p(\alpha) \qquad (11)$$

where $\int d\alpha := \int_0^{2\pi} d\alpha_1 \int_0^{2\pi} d\alpha_2 \ldots \int_0^{2\pi} d\alpha_N$. In ref. [18], from the equivalence of the two averages eq. (10) and eq. (11) for any property $f_P(\alpha)$, it has been shown that, if the energy levels are rationally independent, then the phases are homogeneously distributed, that is

$$p(\alpha) = 1/(2\pi)^N \qquad (12)$$

In conclusion such a probability density, together with the condition of constant populations, defines the Pure State Distribution (PSD), which is the distribution on parameters induced by the time dependence of the quantum pure state. With the aid of the PSD, the time average of any property of a quantum pure state can be easily calculated.

Standard observables of quantum system are provided by time dependent expectation values $a(t)$ of operators $A$

$$a(t) := \langle \psi(t) | A | \psi(t) \rangle = \mathrm{Tr}\{A\rho(t)\} \qquad (13)$$

where $\rho(t)$ is the pure state density matrix operator

$$\rho(t) := |\psi(t)\rangle\langle\psi(t)| = \sum_{k,k'=1}^{N} \sqrt{P_k P_{k'}} \exp\{-i\alpha_k(t) + i\alpha_{k'}(t)\} |E_k\rangle\langle E_{k'}| \qquad (14)$$

with constant diagonal elements

$$\rho_{k,k} = P_k \qquad (15)$$

and oscillating off-diagonal elements

$$\rho_{k,k'}(t) = \sqrt{P_k P_{k'}} \exp\{-i\alpha_k(0) + i\alpha_{k'}(0) - i(E_k - E_{k'})t/\hbar\} \qquad (16)$$

for $k,k' \leq N$, while $\rho_{k,k'}(t) = 0$ for $k > N$ and/or $k' > N$. As long as the observable of eq. (13) includes many oscillating contributions due to the off-diagonal elements of $\rho(t)$, a fluctuating behaviour is expected for $a(t)$. The equilibrium value of the observable is then identified with its asymptotic time average

$$\bar{a} = \lim_{T \to \infty} \frac{1}{T} \int_0^T dt \, a(t) = \text{Tr}\{A\bar{\rho}\} \tag{17}$$

where $\bar{\rho}$ is the time averaged of the density matrix eq. (14), which can be evaluated by employing the phase average eq. (11) with the PSD eq. (12)

$$\bar{\rho} = \sum_{k=1}^{N} P_k |E_k\rangle\langle E_k| \tag{18}$$

The equilibrium average of $a(t)$ is simply given as

$$\bar{a} = \sum_{k=1}^{N} P_k A_{k,k} \tag{19}$$

with an explicit dependence on the populations. It should be emphasized that in the same way one can evaluate the amplitude of fluctuations. Let us introduce the deviation $\Delta a(t)$ of the property $a(t)$ from its average $\bar{a}$

$$\Delta a(t) := a(t) - \bar{a} = \text{Tr}\{A[\rho(t) - \bar{\rho}]\} \tag{20}$$

Then the amplitude of fluctuations is quantified by the time average $\overline{|\Delta a|^2} = \overline{\Delta a \Delta a^*}$ for the general case of complex observable $a(t)$ in correspondence of an operator $A$ which is not self-adjoint. By performing the average by means of the phase integration with PSD, one obtains

$$\overline{|\Delta a|^2} = \sum_{k \neq k'} P_k P_{k'} |A_{k,k'}|^2 \tag{21}$$

with, again, an explicit dependence on the populations.

As specific observables we consider the elements of the reduced density matrix $\mu(t)$ of a subsystem. Let us partition the overall isolated system into a subsystem of interest and its environment. Correspondingly, the overall Hilbert space $\mathcal{H}$ is factorized into the Hilbert subspace $\mathcal{H}^{sub}$ for the subsystem and the Hilbert subspace $\mathcal{H}^{env}$ for the environment:

$$\mathcal{H} = \mathcal{H}^{sub} \otimes \mathcal{H}^{env}. \tag{22}$$

The reduced density matrix for the subsystem is then derived by tracing out the complete density matrix $\rho(t)$ over the environment states

$$\mu(t) := \text{Tr}_{env}\{\rho(t)\} \tag{23}$$

For a given orthonormal basis of subsystem states,

$$|p\rangle, |p'\rangle \in \mathcal{H}_{sub}: \quad \langle p|p'\rangle = \delta_{p,p'} \tag{24}$$

the elements $\mu_{p,p'}(t)$ of the reduced density matrix can be calculated according to eq.(13) by considering the expectation value of the operator

$$A = \left(|p'\rangle\langle p|\right) \otimes 1_{en} \tag{25}$$

where $1_{env}$ is the unity operator in $\mathcal{H}_{env}$. Notice the asymptotic time average of the reduced density matrix

$$\bar{\mu} = \text{Tr}_{env}\{\bar{\rho}\} \tag{26}$$

can be evaluated directly from eq. (18).

In order to illustrate the application of the previous treatment, we shall consider a set of $n$ Randomly Perturbed Einstein Oscillators (RPEO). The reference system is the Einstein model of a solid in which each atom vibrates independently with constant frequency, say $\omega_0$, in the potential well of its neighbors' force fields. In the zero order Hamiltonian $H^0$ we include the independent contribution of each oscillator

$$H^0 = \sum_{j=1}^{n} H_{EO}^{(j)} \tag{27}$$

where $H_{EO} = \sum_{m=1}^{\infty} |e_m\rangle e_m \langle e_m|$ is the Hamiltonian of an harmonic oscillator having $|e_m\rangle$ as eigenstates for $m = 0, 1, 2, \cdots$, and $e_m = \hbar\omega_0(m+1/2)$ as eigenenergies. The eigenvalue problem for the zero order Hamiltonian has the obvious solution

$$H^0 \left|E_M^0\right\rangle = E_M^0 \left|E_M^0\right\rangle \tag{28}$$

where $\left|E_M^0\right\rangle=\left|e_{m_1}\right\rangle\left|e_{m_2}\right\rangle\cdots\left|e_{m_n}\right\rangle$, $E_M^0=\sum_{j=1}^n\hbar\omega_0(m_j+1/2)$, with the set of indices $M=(m_1,m_2,\cdots,m_n)$ describing the eigenstate of the ensemble of oscillators. The active Hilbert space $\mathcal{H}_N$ for a given cutoff energy $E_{\max}$ is identified by imposing the constraint eq. (3) to the zero order energies, that is by including in $\mathcal{H}_N$ the components $\left|E_M^0\right\rangle$ with $E_M^0<E_{\max}$. Furthermore we add to the Hamiltonian a random perturbation contribution $H^1$ meant to take into account generic interactions amongst the oscillators modeled through a $N\times N$ Gaussian Orthogonal Random Matrix (GORM) $W_{GORM}$ in the $\left|E_M^0\right\rangle$ representation. Such a matrix is a realization of the Gaussian Orthogonal Ensemble [30] (GOE) which is completely characterized by its probability density

$$p_{GOE}(W_{GROM})=C\exp\left(-\frac{1}{2\sigma_W^2}Tr\{W_{GROM}^2\}\right) \tag{29}$$

where $\sigma_W$ is the variance within the ensemble of the off-diagonal elements of the matrix. The independent elements of a Gaussian Orthogonal Random Matrix are Gaussian random numbers with the following statistical properties

$$\left\langle W_{ij}\right\rangle_{GOE}=0 \qquad \left\langle W_{ij}^2\right\rangle_{GOE}=\frac{\sigma_W^2}{2}\left(1+\delta_{ij}\right) \tag{30}$$

where $\left\langle\cdots\right\rangle_{GOE}$ is the average with respect to the distribution eq. (29).

In our calculation we set $\sigma_W=1$, while the interaction Hamiltonian is defined as

$$H^1=\lambda W_{GROM} \tag{31}$$

where $\lambda$ is a control parameter assuring that $H^1$ acts like a small perturbation to $H^0$, that is $\left|H^1\right|<<\left|H^0\right|$. It should be stressed that $H^1$ not only eliminates the degeneracy of the zero order energies $E_M^0$, but also ensures the condition of rationally independence of eigenenergies as invoked in the derivation of the Pure State Distribution eq. (12).

In conclusion the Randomly Perturbed Einstein Oscillator model is characterized by three parameters, the number $n$ and the frequency $\omega_0$ of oscillators and the strength of the random perturbation, to be specified in any calculation together with the cutoff energy $E_{max}$. The number of independent parameters can be reduced by using $\hbar\omega_0$ as the energy unit.

From the numerical diagonalization of the complete Hamiltonian $H = H^0 + H^1$ represented on the $|E_M^0\rangle$ bases, one gets both the exact eigenenergies $E_k$ and the corresponding principal directions $|E_k\rangle$. In order to evaluate the time dependent wavefunction $|\psi(t)\rangle$, we have still to choose its initial state $|\psi(0)\rangle$. We employ a random initial state [31] parameterized as

$$|\psi(0)\rangle = \sum_{k=1}^{N} \xi_k |E_k\rangle \bigg/ \sqrt{\sum_{k'=1}^{N} |\xi_{k'}|^2} \qquad (32)$$

where $\xi_k$ for $k = 1, 2, \cdots, N$ are a realization of a set of $N$ independent random complex variables with a Gaussian probability distribution characterized by vanishing average and unitary variance of variables $\xi_k$.

This simple model, as long as the dimension $N$ of the active space $\mathcal{H}_N$ is not exceedingly large, allows the direct calculation of the time evolution of an observable $a(t)$ eq. (13) displaying a non trivial behavior. To illustrate it, in Figure 1 we have represented a portion of the time evolution for the first diagonal element $\mu_{0,0}(t)$ of the reduced density matrix of an oscillator, in correspondence of the model parameters specified in the Figure captions. Panels A and B of the Figure refers to two different random choices of the initial wavefunction. It is evident that the time evolution of $\mu_{0,0}(t)$ has a random character which can be rationalized in terms of fluctuations around the average $\bar{\mu}_{0,0}$ (displayed in the Figure as a red dotted line). Such a behavior is a direct consequence of the superposition of a large number of oscillating contributions brought by the off-diagonal elements of the density matrix eq. (14). By sampling $\mu_{0,0}(t)$ with respect to time $t$, one can obtain its statistical distribution as shown in panels C and D. From this distribution one gets information not only on the average, but also on the amplitude of fluctuations. It is evident that for the two particular cases reported in Fig.1, one derives distributions differing mainly for the shift of their maximum corresponding to the average $\bar{\mu}_{0,0}$. This is a direct consequence of the difference on the populations in the two cases due to the random choices of the initial wavefunction.

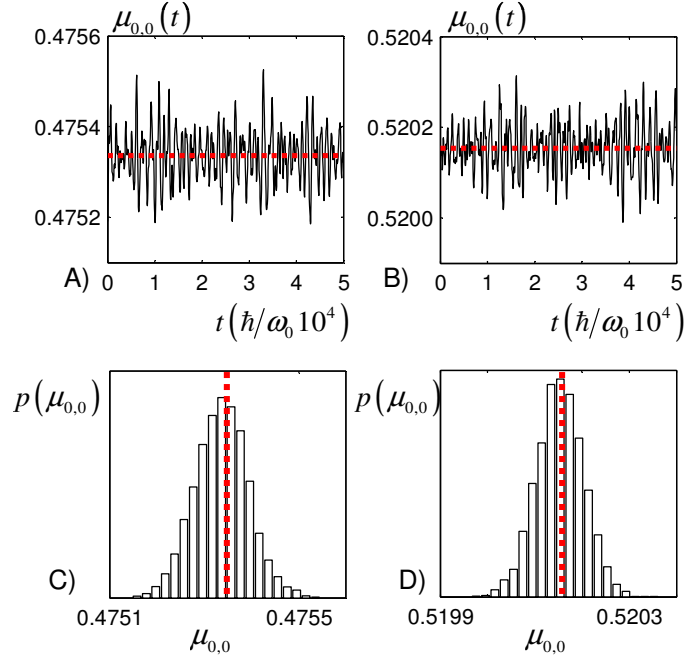

**Figure 1: Time evolution and distribution of the first diagonal element $\mu_{0,0}$ of the reduced density matrix for an oscillator in the Randomly Perturbed Einstein Oscillator model. The following parameters have been employed: $n = 5$, $E_{max} / \hbar \omega_0 = 5.1$ (corresponding to a dimension $N = 252$ of the active Hilbert space), $\lambda / \hbar \omega_0 = 10^{-3}$. Panels A and B display the evolution $\mu_{0,0}(t)$ for two different random choices of the initial wavefunction, with the corresponding statistical distributions reported in panels C and D. The asymptotic time average is indicated by the red dotted line.**

As a further category of observables, we consider also the properties of the overall isolated system in relation to its thermodynamical characterization. They include the microscopic internal energy $\hat{U}$ identified with the expectation value of the Hamiltonian

$$\hat{U} := \langle \psi(t) | H | \psi(t) \rangle = \sum_{k=1}^{N} P_k E_k \tag{33}$$

and the microscopic entropy $\hat{S}$ for the Shannon entropy of the populations

$$\hat{S} := -k_B \sum_{k=1}^{N} P_k \ln P_k \qquad (34)$$

which quantifies the statistical disorder (or lack of information) with respect to the eigenenergy decomposition of the wavefunction (a vanishing entropy is recovered for a stationary state, that is for $|\psi(t)\rangle = |E_m\rangle \exp\{-i\alpha_m(t)\}$ in corresponding of a given eigenstate $|E_m\rangle$). These microscopically defined properties are not strictly equivalent to macroscopic properties. The latter, being thermodynamic state properties, cannot depend on the microscopic details of a quantum system like the populations in eq. (33) and eq. (34). The connections between these microscopic properties and the corresponding thermodynamical state properties will be examined in section IV.

Within the general description supplied by the Pure State Distribution of an isolated quantum system, one can wonder where the standard microcanonical statistics has to be placed. The key ingredient is the statistical density matrix employed in the microcanonical statistics which could be identified with the asymptotic time average eq. (18) for the overall pure state density matrix. However, such an averaged density matrix depends on the constants of motion of the Schrödinger dynamics, that is the populations, and in order to recover microcanonical statistics one has to choose them in a particular way: identical non vanishing populations within a tiny shell of energy eigenstates between two boundaries $E_{\min}$ and $E_{\max}$. It should be evident that in general it is impossible to fix in advance the populations of an isolated quantum system. As long as a particular choice of the populations is arbitrary, it should be preferable to treat them as free parameters which can assume different values depending on the realization of the isolated quantum system (see the next section).

To conclude this section, we emphasize that PSD allows the description not only of time averaged properties, but also of the amplitude of their fluctuations according to eq. (21). Such a feature is completely absent in standard quantum statistical mechanics and, because of the strict relation between fluctuations and relaxation, the analysis of the fluctuations is left to a future contribution focusing on the description of relaxation phenomena in isolated quantum systems. In the following sections we shall confine our analysis to equilibrium average properties.

## III. Random Pure State Ensemble (RPSE)

If we confine our interest to equilibrium properties of an isolated system, then we have to determine functions $f(P)$ of populations $P = (P_1, P_2, \cdots, P_N)$ satisfying the constraint eq. (9). This is the case i) of the time average $\overline{a}$ eq. (19) for the expectation value of a generic operator $A$ like that of eq. (25) for the elements of the reduced density matrix of a subsystem, ii) of the microscopic internal energy $\hat{U}$ eq. (33) and iii) of the microscopy entropy $\hat{S}$ eq. (34). As long as these functions are well defined, the assignment of equilibrium properties to a specific system becomes a trivial operation once the populations are given. However, the knowledge of the population set is far from being accessible. As a matter of fact, it is impossible to prepare a system with several degrees of freedom in an initial state $\psi(0)$ with a specific set of populations. On the other hand, one cannot recover the full set of populations from the limited amount of information supplied by a measurement. In conclusion a full knowledge of the populations of a given isolated system is behind our means.

This is the typical framework for the statistical mechanics. Therefore one has to resort to a statistical analysis of equilibrium properties on the basis of the probabilistic distribution of the populations. For this purpose we employ the concept of statistical ensemble [22] specified through the sample space $D$ of the possible population sets, and the corresponding probability density $p(P)$. In the $N$-th dimensional space for the populations $(P_1, P_2, \cdots, P_N)$ considered as independent parameters, the sample space is the (N-1)-th dimensional simplex deriving from the population normalization eq. (9) and from the positivity of each population (see Fig.1 of ref. [18] for an illustration of the sample space)

$$D = \left\{ (P_1, P_2, \cdots, P_N) \in \mathbb{R}^N \mid \sum_{k=1}^{N} P_k = 1, P_k \geq 0 \, \forall k \right\} \tag{35}$$

In principle different ensembles can be introduced on the basis of particular choices for the probability density $p(P)$ on the populations. In the present work we shall employ the so-called

Random Pure State Ensemble (RPSE), which corresponds to an initial pure state $\psi(0)$ randomly chosen according to the uniform distribution on the unit sphere in the active space Hilbert $\mathcal{H}_N$.[32]

By considering $(P_1, P_2, \cdots, P_{N-1})$ as the set of independent populations, from a geometrical analysis of the measure in the Hilbert space [33] one derives a constant probability density for the RPSE

$$p(P_1, P_2, \cdots, P_{N-1}) = (N-1)! \tag{36}$$

with normalization

$$\int dP_1 dP_2 \cdots dP_{N-1} p(P_1, P_2, \cdots, P_{N-1}) = 1 \tag{37}$$

where the integration domain is determined by the sample space $D$. With such an explicit form of the probability density, one in principle can calculate the RPSE average of any function $f(P_1, P_2, \cdots, P_{N-1})$ of the populations, which in the following will be denoted by means of a bracket:

$$\langle f \rangle := \int dP_1 dP_2 \cdots dP_{N-1} p(P_1, P_2, \cdots, P_{N-1}) f(P_1, P_2, \cdots, P_{N-1}) \tag{38}$$

It should be explicitly noted that the $(N-1)$ relevant populations are not statistically independent since the existence domain of each population depends on the others. Let us specify the order of integration as $dP = dP_1 dP_2 ... dP_{N-1}$. Then the condition of positivity of the last population, $P_N = 1 - \sum_{i=1}^{N-1} P_i > 0$, determines the allowed region for $P_{N-1}$ as a function of the previous populations, $P_{N-1} \leq b_{N-2}(P_1, ..., P_{N-2}) := 1 - \sum_{j=1}^{N-2} P_j$. Furthermore, by requiring the upper bound $b_{N-2}$ to be a positive number, one finds the upper bound for the population $P_{N-2}$, and so on. Since the integral on populations $P_{J+1}, P_{J+2}, \cdots, P_{N-1}$ can be solved analytically, one can obtain the exact joint probability density on $J$ populations of the Random Pure State Ensemble

$$p(P_1, \cdots, P_J) = \int_0^{b_J} dP_{J+1} ... \int_0^{b_{N-2}} dP_{N-1} p(P_1, ... P_{N-1}) = \frac{(N-1)!}{(N-J-1)!} b_J(P_1, \cdots, P_J)^{N-J-1} \tag{39}$$

where $b_J(P_1,...P_J) := 1 - \sum_{j=1}^{J} P_j$. By choosing $J=1$, one derives the distribution function on the first population

$$p(P_1) = (N-1)(1-P_1)^{N-2} \tag{40}$$

An equivalent result is obtained for the marginal distribution of any population, because of the invariance of the full distribution eq. (36) with respect to the population exchange. Then the first two moments of a population are easily obtained by integration with eq. (40)

$$\langle P_k \rangle = 1/N \qquad \langle P_k^2 \rangle = 2/N(N+1) \tag{41}$$

In the following analysis we need also the average of the product between two populations, $\langle P_k P_{k'} \rangle$ for $k \neq k'$. From the distribution on the first two populations,

$$p(P_1, P_2) = (N-1)(N-2)(1-P_1-P_2)^{N-3} \tag{42}$$

and again by taking into account the invariance with respect to population exchange, we obtain

$$k \neq k': \qquad \langle P_k P_{k'} \rangle = 1/N(N+1) = \langle P_k^2 \rangle / 2 \tag{43}$$

In order to determine the distribution of any property within the RPSE, it is convenient to get a sample of the population sets within such an ensemble. This can be easily done numerically by means of an auxiliary set of $(N-1)$ independent random variables $\xi \equiv (\xi_1,...,\xi_{N-1})$ uniformly distributed in $(0,1]$. It is easily shown [33,34] that the set of populations calculated as

$$P_1 = 1 - \xi_1^{\frac{1}{N-1}}, \quad ......, \quad P_J = \left(1 - \xi_J^{\frac{1}{N-J}}\right) \prod_{i=1}^{J-1} \xi_i^{\frac{1}{N-i}}, \quad ......, \quad P_N = \prod_{i=1}^{N-1} \xi_i^{\frac{1}{N-i}} \tag{44}$$

is a realization from the RPSE distribution. As an example, in Fig. 2 we have reported the distribution for different population realizations of the RPSE, of the equilibrium average $\bar{\mu}_{0,0}$ of the ground state diagonal element of the reduced density matrix of a single oscillator, within the Randomly Perturbed Einstein Oscillator model previously introduced. Notice that, despite a similar appearance, the distributions reported in Fig. 1 and in Fig. 2 describe completely different situations. While the distributions of panels C and D of Fig. 1 are determined, for a given isolated system, by the time evolution of the instantaneous reduced density matrix $\mu(t)$ which fluctuates around its equilibrium average $\bar{\mu}$, on the contrary the distributions of Fig. 2 describe the possible values of the equilibrium property as recovered by different realizations (i.e., the different possible population sets) of the isolated system. In other words, each panels of Fig. 1 refers to a given isolated systems, while Fig. 2 describes the statistics for a collection of the same type of isolated systems.

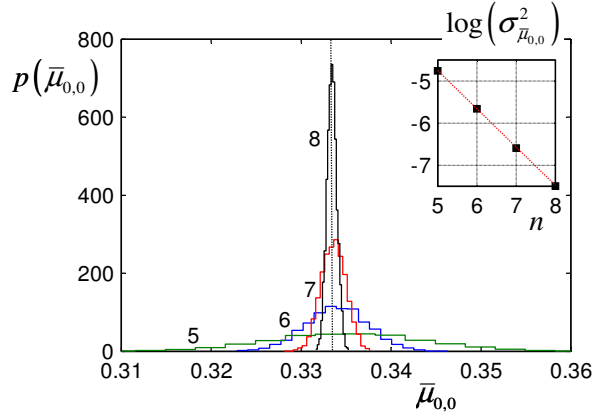

**Figure 2 Distribution within RPSE of first element, $\bar{\mu}_{0,0}$, of the equilibrium reduced density matrix of a single oscillator of the RPEO model as obtained from the sampling of population sets. The distributions refers to systems composed of different number of oscillators (n=5,6,7,8), with $E_{\max}/n\hbar\omega_0 = 2$ and $\lambda/\hbar\omega_0 = 10^{-3}$. In the inset we have reported the numerically determined variance $\sigma_{\bar{\mu}_{0,0}}$ (in a logarithmic scale) as a function of the number of oscillators.**

Distributions like those of Fig. 2 are obtained for other equilibrium properties $f(P)$ considered as functions of the populations. In the particular case of $f = \bar{\mu}_{0,0}$, Fig. 2 provides the evidence that such a property is characterized by a clearly peaked distribution with widths much smaller than the range $\Delta_f$ for the possible values of the property ($\Delta_f = 1$ for $f = \bar{\mu}_{0,0}$, since the minimum and the maximum values of $\mu_{0,0}$ are 0 and 1, respectively). The essential features of the distribution are determined by the mean value $\langle f \rangle$ and the variance

$$\sigma_f := \sqrt{\langle (f - \langle f \rangle)^2 \rangle} \tag{45}$$

both calculated with respect to the RPSE statistics. The same fact that the variance is much smaller than the range of possible values

$$\sigma_f \ll \Delta_f \tag{46}$$

implies that property $f$ manifests typicality.

The concept of typicality has its origin in the field of information theory [35], but in recent years has found a widespread use in statistical mechanics [11,12,13,14,15]. A generic form of typicality is associated to an event $A$ whose probability $\text{Prob}(A)$ is very close to unity:

$$\text{Prob}(A) = 1 - \varepsilon \tag{47}$$

for a small $\varepsilon$. In other words the exceptions are very unlikely since the probability of the complementary event $\bar{A}$ is small, $\text{Prob}(\bar{A}) = \varepsilon \ll 1$. For instance in our case, a nearly unitary probability (more precisely about 95%) is found for the event that the property is within $\langle f \rangle - 2\sigma_f \leq f \leq \langle f \rangle + 2\sigma_f$, that is within an interval much smaller than the possible values of the property. A stronger form of typicality is assigned to an almost sure event $A$ with a unitary probability

$$\text{Prob}(A) = 1 \tag{48}$$

The property $f = \bar{\mu}_{0,0}$ displays such a form of typicality because, as shown in the inset of Fig. 3, the variance decreases very quickly (as a matter of fact, exponentially) with the number of oscillators: $\lim_{n \to \infty} \sigma_f / \Delta_f = 0$. Therefore the probability the actual value of the property $f$ is within a given interval centered on its average, $\langle f \rangle - a \leq f \leq \langle f \rangle + a$ for fixed $a$, tends to unity in the limit of an infinite number of oscillators. Similarly, typicality in its strong form, eq. (48), holds for any property $f$ as long as $\sigma_f / \Delta_f$ tends to vanish in a suitable limit condition. Often such a feature is described by stating that the property $f$ is "almost surely" equal to its average $\langle f \rangle$.

Typicality has important implications on the description of equilibrium properties according to statistical ensemble on the populations. Even if on the basis of reasonable arguments one can assume a well defined statistics, for instance RPSE, this in general does not provide a definite answer about the value of an equilibrium property in an actual system. Because of the population dependence $f(P)$ of an equilibrium property, one can derive only the distribution of its possible values, without the possibility of assigning to it a well defined value. However, if the strong form of typicality is assured in a proper limit condition, then one can attribute to $f$ almost surely the value $\langle f \rangle$. In conclusion, definite prediction about the actual value of the property $f$ can be derived, even if $f$ depends on the populations.

In the rest of this section we will demonstrate the typicality of the relevant equilibrium properties $f$ for an isolated quantum system in the limit of infinite dimension $N$ of the active space $\mathcal{H}_N$, by verifying the validity of the limits

$$\lim_{N \to \infty} \sigma_f / \Delta_f = 0 \tag{49}$$

It should be emphasized that these demonstrations are very general since only the asymptotic condition on the dimension $N$ is required, without any specific requirement about the nature of the quantum system. The meaning of such a formal limit in relation with the thermodynamic limit will be clarified in the next section.

Let us first examine the microscopic internal energy $\hat{U}$ defined by eq. (33). Its RPSE average, taking into account that according to eq. (41) the averages of all the populations are equal to $1/N$, is given as

$$\langle \hat{U} \rangle = \sum_{k=1}^{N} E_k / N \tag{50}$$

We introduce the density of states in the full Hilbert space $\mathcal{H}$

$$G(E) := \sum_{k=1}^{\infty} \delta(E - E_k) \tag{51}$$

such that the dimension $N$ of the active space $\mathcal{H}_N$ is recovered as the number of states with energy up to the cut-off $E_{max}$

$$N = \int_{-\infty}^{E_{max}} G(E) dE \tag{52}$$

Then the average of the internal energy can be rewritten as

$$\langle \hat{U} \rangle = \int_{-\infty}^{E_{\max}} E \frac{G(E)}{N} dE \qquad (53)$$

and it can be interpreted as the average energy deriving from the normalized density of states $G(E)/N$ for a given cutoff $E_{\max}$. The variance of the microscopic internal energy

$$\sigma_{\hat{U}}^2 = \left\langle \left( \hat{U} - \langle \hat{U} \rangle \right)^2 \right\rangle = \left\langle \left[ \sum_{k=1}^{N} \left( E_k - \langle \hat{U} \rangle \right) P_k \right]^2 \right\rangle \qquad (54)$$

by evaluating the moments of the populations according to eq. (41) and to eq. (43), is given as

$$\sigma_U^2 = \frac{1}{N(N+1)} \sum_{k=1}^{N} \left( E_k - \langle \hat{U} \rangle \right)^2 = \frac{1}{N+1} \int_{-\infty}^{E_{\max}} \left( E - \langle \hat{U} \rangle \right)^2 \frac{G(E)}{N} dE \qquad (55)$$

Let us introduce the parameter $\kappa$

$$\kappa^2 := \frac{1}{\Delta_{\hat{U}}^2} \int_{-\infty}^{E_{\max}} \left( E - \langle \hat{U} \rangle \right)^2 \frac{G(E)}{N} dE \qquad (56)$$

which describes the ratio between the mean squared energy deviations computed from the density of states, and the squared range $\Delta_{\hat{U}} := E_{\max} - E_1$ of the possible value of the energy (and of the internal energy as well). It should be noticed that $\kappa$ is of the order of unity if the density of state weakly depends on the energy (in particular $\kappa = 1/2\sqrt{3}$ if the density of states is constant), while $\kappa \ll 1$ in the opposite situation of a fast increasing density of states. Then one derives the following relation for the variance

$$\sigma_{\hat{U}} / \Delta_{\hat{U}} = \kappa / \sqrt{N+1} \qquad (57)$$

which, according to eq. (49), implies typicality for the internal energy independently of the parameter $\kappa$ determined by the profile of the density of states.

The RPSE statistical properties of the microscopic entropy are derived on the basis of its definition eq. (34). Its average, taking into account the invariance with respect to population exchange, can be specified as

$$\langle \hat{S} \rangle = -Nk_B \langle P_1 \ln P_1 \rangle \qquad (58)$$

In order to evaluate its limit for $N \to \infty$, one must consider that the typical values of the populations shift to zero because $\langle P_k \rangle = 1/N$. Therefore it is convenient to introduce scaled populations

$$x_k := P_k N \tag{59}$$

having constant averages, $\langle x_k \rangle = 1$, so that their typical values remain constant in the limit $N \to \infty$. Correspondingly the average entropy becomes

$$\langle \hat{S} \rangle = k_B \ln N - k_B \langle x_1 \ln x_1 \rangle \tag{60}$$

In Appendix A it is shown that in the limit $N \to \infty$ the RPSE average $\langle g(x_k) \rangle = \langle g(x_1) \rangle$ of any function $g(x_k)$ of a single scaled population can be evaluated as

$$N \to \infty: \quad \langle g(x_1) \rangle = \langle g(x_1) \rangle_\infty - \frac{1}{N} \langle g(x_1)(1 - 2x_1 + x_1^2/2) \rangle_\infty \tag{61}$$

where the first term

$$\langle g(x_1) \rangle_{N \to \infty} := \lim_{N \to \infty} \langle g(x_1) \rangle = \int_0^\infty dx_1 g(x_1) e^{-x_1} \tag{62}$$

is the leading contribution, while the second term represents its first order correction with respect to $1/N$. This implies that in eq. (60) $k_B \ln N$ is the leading contribution in the limit $N \to \infty$ for the RPSE average of the entropy

$$N \to \infty: \quad \langle \hat{S} \rangle = k_B \ln N \tag{63}$$

From the deviation of the microscopic entropy from its average, written in terms of the scaled populations

$$\hat{S} - \langle \hat{S} \rangle = -\frac{k_B}{N} \sum_{k=1}^N \left( x_k \ln x_k - \langle x_k \ln x_k \rangle \right) \tag{64}$$

one derives the following relation for its RPSE variance:

$$\sigma_{\hat{S}}^2 = \left\langle \left( \hat{S} - \langle \hat{S} \rangle \right)^2 \right\rangle = \frac{k_B^2}{N} \left[ \langle (x_1 \ln x_1)^2 \rangle - \langle x_1 \ln x_1 \rangle^2 \right] + \\ + k_B^2 \left( 1 - \frac{1}{N} \right) \left[ \langle (x_1 \ln x_1)(x_2 \ln x_2) \rangle - \langle x_1 \ln x_1 \rangle \langle x_2 \ln x_2 \rangle \right] \tag{65}$$

As shown in Appendix A, the following relation hold for the generalization of eq. (61) for the RPSE average of the product of the same function of two different populations

$$N \to \infty : \langle g(x_1)g(x_2)\rangle = \langle g(x_1)\rangle_\infty^2 - \frac{1}{N}\left\{\langle x_1 g(x_1)\rangle_\infty^2 + \langle g(x_1)\rangle_\infty \langle g(x_1)(3 - 6x_1 + x_1^2)\rangle_\infty\right\} \tag{66}$$

By substitution the expansions eq. (61) and eq. (66) into eq. (65), one derives that the zero-order contribution with respect to $1/N$ vanishes, and the following relation is obtained for the leading contribution to the microscopic entropy variance

$$N \to \infty : \quad \sigma_{\hat{S}}^2 = \frac{k_B^2}{N}\left[\langle (x_1 \ln x_1)^2\rangle_\infty - \langle x_1^2 \ln x_1\rangle_\infty^2 - 2\langle x_1 \ln x_1\rangle_\infty \langle (1 - x_1)x_1 \ln x_1\rangle_\infty\right] \tag{67}$$

or, by employing the explicit values of the integrals which are reported in Appendix A

$$N \to \infty : \quad \sigma_{\hat{S}}^2 = \frac{k_B^2}{N}\left(\frac{\pi^2}{3} - 3\right) \tag{68}$$

In conclusion the $1/\sqrt{N}$ asymptotic dependence is found for the variance of the microscopic entropy and this result assures its typicality according to eq. (49), by taking into account that $0 \leq \hat{S} \leq \Delta_{\hat{S}} = k_B \ln N$.[36]

In order to analyze the typicality of the subsystem reduced density matrix, once the full Hilbert space is factorized according to eq. (22), the full Hamiltonian can be decomposed as

$$H = H^{sub} \otimes 1^{env} + 1^{sub} \otimes H^{env} + H^{sub,env} \tag{69}$$

where $H^{sub}$ and $H^{env}$ are the Hamiltonians for the subsystem and the environment, respectively, with the corresponding eigenvalue problems specified as

$$H^{sub}|E_m^{sub}\rangle = E_m^{sub}|E_m^{sub}\rangle \qquad H^{env}|E_j^{env}\rangle = E_j^{env}|E_j^{env}\rangle \tag{70}$$

while $H^{sub,env}$ represents the interaction between subsystem and environment. Furthermore we suppose that contributions of the interaction term $H^{sub,env}$ are small enough to be considered as negligible. Such a condition is justified if we are free to choose the subsystem as a large enough part of the overall isolated system, as long as subsystem-environment interactions scale as the subsystem surface, while the subsystem or environment energies scale according to their volume.

In such a case the eigenvalues and the eigenvectors of the overall system are well approximated according to the contributions of the subsystem and of the environment alone:

$$E_k = E_m^{sub} + E_j^{env} \qquad |E_k\rangle = |E_m^{sub}\rangle|E_j^{env}\rangle \qquad (71)$$

with $k = (m, j)$. Correspondingly the populations are denoted as $P_k = P_{m,j}$, and they are not vanishing only for

$$E_m^{sub} + E_j^{env} < E_{max} \qquad (72)$$

$E_{max}$ being the energy cut-off which defines the active Hilbert space eq.(3). According to eqs. (26) and (18), the averaged reduced density matrix $\bar{\mu}$ is diagonal in such a basis, with diagonal elements

$$\bar{\mu}_{m,m} = \sum_j P_{m,j} \theta(E_{max} - E_m^{sub} - E_j^{env}) \qquad (73)$$

where the step function, $\theta(x) = 1$ for $x > 0$ and $\theta(x) = 0$ otherwise, enforces the condition (72). Then the RPSE average of the reduced density matrix elements are given as

$$\langle \bar{\mu}_{m,m} \rangle = (1/N) \sum_j \theta(E_{max} - E_m^{sub} - E_j^{env}) \qquad (74)$$

since the population RPSE average are all equal to $1/N$. The corresponding variance

$$\sigma_{\bar{\mu}_{m,m}}^2 = \left\langle \left(\bar{\mu}_{m,m} - \langle \bar{\mu}_{m,m}\rangle\right)^2 \right\rangle = \langle \bar{\mu}_{m,m}^2 \rangle - \langle \bar{\mu}_{m,m} \rangle^2 \qquad (75)$$

requires the calculation of the RPSE average of the square of eq. (73)

$$\langle \bar{\mu}_{m,m}^2 \rangle = \sum_j \theta(E_{max} - E_m^{sub} - E_j^{env}) \sum_{j'} \theta(E_{max} - E_m^{sub} - E_{j'}^{env}) \langle P_{m,j} P_{m',j} \rangle \qquad (76)$$

By using eqs. (41) and (43) for the average of population products and by employing eq. (74) to specify the summation on the unity step function, we get after some algebra the following result for the reduced density matrix RPSE variance

$$\sigma_{\bar{\mu}_{m,m}}^2 = \frac{\langle \bar{\mu}_{m,m}^2 \rangle - \langle \bar{\mu}_{m,m} \rangle^2}{N+1} \qquad (77)$$

that is $\sigma_{\bar{\mu}_{m,m}} \propto 1/\sqrt{N+1}$, which implies typicality according to eq. (49), since $\Delta_{\bar{\mu}_{m,m}} = 1$.

## IV. Thermodynamical behaviour

We now turn to the main issue treated in this work: in order to recover equilibrium thermodynamics from the previous statistical description of quantum systems, we have to identify the internal energy $U$ and the entropy $S$ of macroscopic systems. To simplify the matter we shall concentrate our attention on the thermal properties of systems at constant volume so that the fundamental differential of thermodynamics is reduced to $dU = TdS$. As long as large size systems have to be considered, then nearly infinite values have to be assigned to the dimension $N$ of the active Hilbert space and, therefore, we can safely assume typicality for the microscopic internal energy and entropy in correspondence of their RPSE averages $\langle \hat{U} \rangle$ and $\langle \hat{S} \rangle$. These averages provide the natural definition of the macroscopic (thermodynamic) properties internal energy $U$ and entropy $S$ since, in opposition to their microscopic counterparts, they are independent of the particular realization of the quantum state, that is, they are independent of the population set. Then, in agreement with eq. (50) and eq. (63), the following relations will be employed for the calculation of the thermodynamic internal energy and entropy

$$U := \langle \hat{U} \rangle = \sum_{k=1}^{N} E_k / N \tag{78}$$

$$S := \langle \hat{S} \rangle = k_B \ln N \tag{79}$$

Notice that, for a given system characterized according to its Hamiltonian or equivalently according to the density of states $G(E)$, the dimension of the active space is a function of the cut-off energy: $N = N(E_{max})$, see eq. (52). Thus, both the thermodynamic properties can be derived once the function $N(E_{max})$ is given.[37] It should be also clear that in this framework both the thermodynamic properties has to be considered as functions of the cut-off energy $E_{max}$.

In order to recover a macroscopic description in agreement with standard thermodynamics, three main requirements should be assured:

i) the existence of the state function $S = S(U)$ for the dependence of the entropy on the internal energy;

ii) both the internal energy and the entropy should be extensive properties, such that the temperature derived from the fundamental differential

$$\frac{1}{T} = \frac{dS(U)}{dU} \qquad (80)$$

results to be an intensive parameter;

iii) the temperature should be positive and an increasing function of the internal energy. This, according to eq. (80), implies that the entropy equation of state $S(U)$ must be a convex increasing function of the internal energy.

The first requirement is readily verified from the functional dependence of the internal energy eq. (78) and the entropy eq. (79) on the cut-off energy: $U = f(E_{max})$ and $S = g(E_{max})$. By eliminating their $E_{max}$ dependence as $S = g(f^{-1}(U))$, the equation of state $S = S(U)$ is recovered.

The characterization of both the internal energy and the entropy as extensive properties needs a more complex analysis since one should determine their dependence on the number of components of the system. The system has to be divided into $n$ identical parts (hereafter called components), each of them bringing an independent contribution to thermodynamic properties $U$ and $S$. The simplest case to analyze is that of ideal systems of non interacting molecules which, in this case, identify the components. Correspondingly the Hamiltonian of the overall system is obtained by summing up the independent Hamiltonians of the components. It should be emphasized that such an ideal representation has to be considered as an approximation of real systems which necessarily include interactions between components having a primary role in their relaxation behaviour. Furthermore, in our framework these interactions are required to remove the degeneracy of the energy spectrum of the overall system and to recover the Pure State Distribution (PSD) for the evolution of a single realization of the overall system. The use of ideal models is justified only in the hypothesis that these interactions are small enough to bring a pertubational contribution to the energy of the overall system. Correspondingly they can be neglected in the calculation of the density of states $G(E)$ and of the number of states $N(E_{max})$. This implies that the use of ideal models is justified for the evaluation of equilibrium properties like internal energy and entropy which are determined by the function $N(E_{max})$ as previously shown.

Of course such a description cannot be applied to systems with strong intermolecular interactions (liquids and solids for instance). In these cases the components have to be identified with identical parts of system, large enough so that their interactions (which scale according to the surface of the components) bring to the overall energy a negligible contribution with respect to those of the components (which scale according to their volume). With such an identification of the components, one recovers the methodological similarity with the treatment of ideal systems.

Once the $n$ identical and non-interacting components have been identified in both the so-called ideal and real systems, we evaluate the thermodynamic properties on the basis of the quantum description of each component. We shall follow the same procedure employed in ref. [19] for the analysis of ideal spin systems, but generalizing it to generic systems. Let us characterize each component by its set of energy eigenvalues $(e_0, e_1, e_2, \ldots)$ ordered in magnitude ($e_m \leq e_{m+1}$) and with $e_0 = 0$ in correspondence of the ground state. The energy eigenstate of the overall isolated system can be characterized by means of the set $i = (i_0, i_1, i_2, \ldots)$ of occupation numbers $i_m$ for each energy eigenvalue $e_m$ of the components. These occupation numbers have to satisfy the constraints on the number $n$ of the components

$$\sum_m i_m = n \qquad (81)$$

and on the overall system energy which should be less than the cut-off energy $E_{max}$

$$E := \sum_m e_m i_m < E_{max} \qquad (82)$$

It should be stressed that, for a given cut-off energy, the set of non vanishing occupation numbers is finite. Indeed the last occupied state $e_M$ is determined by the conditions $e_M < E_{max}$ and $e_{M+1} \geq E_{max}$, with vanishing occupation numbers $i_m$ for $m > M$.

By taking into account that a given set of occupation numbers corresponds to $n!/\prod_m i_m!$ different energy eigenstates for the overall system, one determines the dimension $N$ of the active Hilbert space as a function of the number $n$ of components and of the energy cut-off $E_{max}$

$$N(n, E_{max}) = \sum_i \frac{n!}{\prod_m i_m!} \delta_{n, \sum_m i_m} \theta\left(E_{max} - \sum_m e_m i_m\right) \tag{83}$$

where $\sum_i$ denotes summations on all the occupation numbers:

$$\sum_i \cdots := \prod_m \sum_{i_m} \cdots \tag{84}$$

Notice that constraints eq. (81) and eq. (82) are taken into account by means of the Kronecker symbol $\delta$ and the unit step functions $\theta$ with values $\theta(x) = 1$ if $x > 0$, and $\theta(x) = 0$ otherwise. Because of eq. (52), the energy derivative of the number $N$ of states determines the density of states $G(n, E)$ for the system with $n$ components

$$G(n, E) = \frac{\partial N(n, E)}{\partial E} = \sum_i \frac{n!}{\prod_m i_m!} \delta_{n, \sum_m i_m} \delta\left(E - \sum_m e_m i_m\right) \tag{85}$$

where the last term at the r.h.s. is the Dirac delta deriving from the derivative of the unit step function. Correspondingly the number of states can be derived by integration of the density of states

$$N(n, E_{max}) = \int_0^{E_{max}} G(n, E) dE \tag{86}$$

Since negative energy states are excluded on the basis of the assumed energy spectrum of the components, the integration on the density of states is performed only on positive values of the energy. Given the number of states $N(n, E_{max})$, eventually to be calculated according to eq. (86) by means of the density of states, one can derive the entropy from eq. (79)

$$S(n, E_{max}) = k_B \ln N(n, E_{max}) \tag{87}$$

and the internal energy from eq. (78)

$$U(n, E_{max}) = E_{max} - \int_0^{E_{max}} dE \, \frac{N(n, E)}{N(n, E_{max})} \tag{88}$$

Once the thermodynamic limit is defined according the asymptotic values of the number of components, $n \to \infty$, the condition ii) about extensivity can then be reformulated as the requirement that in such a limit the internal energy per component, $U/n$, and the entropy per component, $S/n$, become functions only of an intensive parameter like the cut-off energy per component

$$e_{max} := E_{max}/n \tag{89}$$

Formally, this is equivalent to require that the following two limits

$$u(e_{max}) := \lim_{n \to \infty} \frac{U(n, ne_{max})}{n} \qquad s(e_{max}) := \lim_{n \to \infty} \frac{S(n, ne_{max})}{n} \tag{90}$$

exists and are described by well behaving functions of the $e_{max}$ parameter. As a matter of fact calculations with specific systems like the Randomly Perturbed Einstein Oscillators model (see Fig. 3) show that by increasing the number $n$ of components (i.e., the number of oscillators) the $e_{max}$ dependence of $U/n$ and of $S/n$ tends to asymptotic profiles. Notice that, in agreement with the previous characterization of the components, the contribution of the perturbation Hamiltonian has not been considered. Furthermore, we will show that the two limits of eqs. (90) can be determined analytically in all generality, that is without requirements on the physical structure of the components.

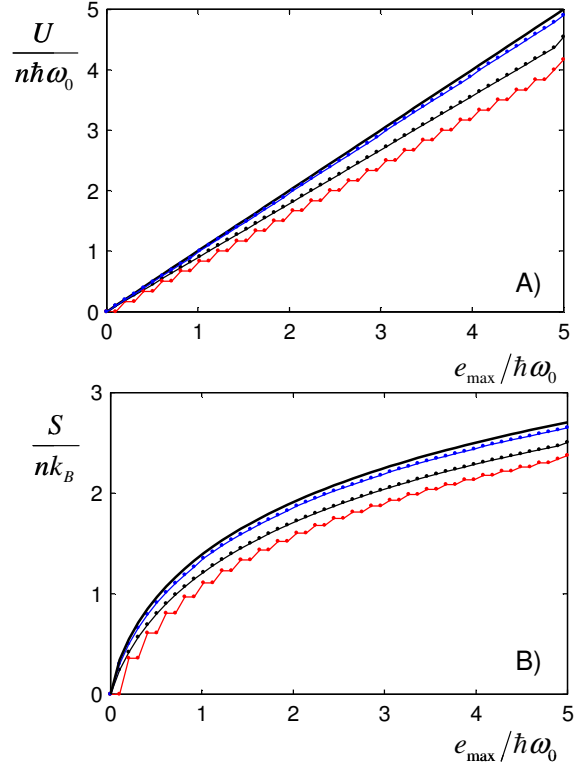

**Figure 3: Scaled internal energy per component** $U/n\hbar\omega_0$ **(panel A) and entropy per component** $S/nk_B$ **(panel B) as a functions of the scaled cut-off energy** $e_{max}/\hbar\omega_0$ **per component, for systems of** $n=5$ **(red points),** $n=10$ **(black points),** $n=50$ **(blue points) oscillators. The asymptotic** $n\to\infty$ **profiles are represented with black continuous lines.**

As shown in detail in Appendix B by analyzing the asymptotic $n\to\infty$ behaviour of the number $N$ of states, the limit eq. (90) for the entropy can be specified as

$$s(e_{max}) = -k_B \sum_m p_m(e_{max}) \ln p_m(e_{max}) \qquad (91)$$

where for any energy level $e_m$ of the components the following parameter is introduced

$$p_m(e_{max}) = \frac{e^{-\beta(e_{max})e_m}}{Q(e_{max})} \qquad Q(e_{max}) = \sum_m e^{-\beta(e_{max})e_m} \qquad (92)$$

with the function $\beta(e_{max})$ given as the $e_{max}$-dependent solution of the equation

$$e_{max} = \sum_m e_m p_m(e_{max}) \qquad (93)$$

These parameters $p_m$, being positive and normalized as

$$\sum_m p_m(e_{\max}) = 1 \tag{94}$$

acquire the meaning of probabilities of the component energy levels $e_m$ and, therefore, according to eq. (93) $e_{\max}$ can be interpreted as the average energy per component.

From these results one can derive that $s(e_{\max})$ is a convex increasing function of $e_{\max}$. By taking into account that the number of states $N(n, E_{\max})$, and the entropy $S(n, E_{\max})$ eq. (87) as well, are increasing functions of $E_{\max}$ for a given number $n$ of components, the condition of positivity is derived for the first derivative of $s(e_{\max})$

$$s'(e_{\max}) := \frac{ds(e_{\max})}{de_{\max}} = \lim_{n \to \infty} \frac{\partial}{\partial e_{\max}} \frac{S(n, ne_{\max})}{n} = \lim_{n \to \infty} \frac{\partial}{\partial E_{\max}} S(n, E_{\max}) \bigg|_{E_{\max} = ne_{\max}} > 0 \tag{95}$$

so demonstrating that $s(e_{\max})$ is an always increasing function. Before to analyze its second derivative, let us rewrite eq. (91) by specifying the probability of component states according to eq.(92)

$$s(e_{\max}) = k_B \ln Q(e_{\max}) + k_B e_{\max} \beta(e_{\max}) \tag{96}$$

Then its derivative, by taking into account that from the second of eqs. (92) $d \ln Q(e_{\max}) / de_{\max} = -e_{\max} \beta'(e_{\max})$ where $\beta'(e_{\max}) := d\beta(e_{\max}) / de_{\max}$, can be specified as

$$s'(e_{\max}) = k_B \beta(e_{\max}) \tag{97}$$

From the derivative of eq. (93) with respect to $e_{\max}$ we obtain

$$1 = -\beta'(e_{\max}) \sum_m p_m(e_{\max})(e_m - e_{\max})^2 \tag{98}$$

where the summation at the r.h.s. represents the variance of the component energy levels $(e_0, e_1, e_2, \ldots)$ computed with probabilities eq. (92) and therefore it is an intrinsically positive quantity. This assures that $\beta'(e_{\max})$ is negative and then, from the derivative of eq. (a19),

$$\frac{d^2 s(e_{\max})}{de_{\max}^2} = k_B \beta'(e_{\max}) < 0 \tag{99}$$

In conclusion, because of the validity of eq. (95) and eq. (99) for all the allowed values $e_{max}$, we have demonstrated that $s(e_{max})$ is a convex increasing function.

Let us now analyze the limit of eqs. (90) for the internal energy. By specifying the number of states according to the entropy eq. (87), and by changing to $e = E/n$ the integration variable, eq. (88) can be rewritten as

$$\frac{U(n, ne_{max})}{n} = e_{max} - \int_0^{e_{max}} de \exp\{-[S(n, ne_{max}) - S(n, ne)]/k_B\} \tag{100}$$

On the other hand for a large number $n$ of components the entropy can be approximated as $S(n, ne) \simeq ns(e)$, and the previous equation becomes:

$$n \to \infty: \quad \frac{U(n, ne_{max})}{n} - e_{max} = -\int_0^{e_{max}} de \exp\{-[s(e_{max}) - s(e)]n/k_B\} \tag{101}$$

We know already from eq. (95) that $s(e)$ is an increasing function of $e$ and therefore the integrand is an increasing function with a slope increasing exponentially with the number of components. This legitimates the substitution of $s(e)$ with its linear expansion about the maximum at $e = e_{max}$, $s(e) = s(e_{max}) + s'(e_{max})(e - e_{max})$, so obtaining the following relation after having shifted to $-\infty$ the lower integration boundary

$$n \to \infty: \quad \frac{U(n, ne_{max})}{n} - e_{max} = -\frac{k_B}{ns'(e_{max})} \tag{102}$$

In conclusion, by substitution into eq. (90), we verify that also the limit for the internal energy exists, and it is precisely the cut-off energy scaled by the number of components

$$u(e_{max}) = e_{max} \tag{103}$$

This implies that in the thermodynamic limit the cut-off energy coincides with the internal energy, $U = E_{max}$, and that the functional dependence of the entropy can be specified as

$$n \to \infty: \quad S = ns(E_{max}/n) = ns(U/n) \tag{104}$$

In the same conditions, the following relation is recovered for the temperature

$$n \to \infty: \quad \frac{1}{T} = \frac{dS}{dU} = \frac{dS}{dE_{max}} = s'(e_{max}) = k_B \beta(e_{max}) \tag{105}$$

with eq. (97) employed to derive the r.h.s.. This result, by substitution into eq. (92), allows us to derive the explicit form of the probabilities for the component states at a given temperature

$$p_m = \frac{e^{-e_m/k_B T}}{\sum_{m'} e^{-e_{m'}/k_B T}} \tag{106}$$

which is the canonical distribution. Besides, as long as the temperature can substitute $e_{max} = U/n$ as the independent variable for the entropy per component, $s = s(T)$, then one finds from eq. (87) that in the thermodynamic limit the overall number of states at fixed temperature increase exponentially with the number of components:

$$n \to \infty: \quad N = \exp\{ns(T)/k_B\} \tag{107}$$

On the other hand, since we have already shown that $s(e_{max})$ is a convex increasing function, also the fundamental requirement iii) is assured. In conclusion in the limit of an infinite number of components we recover for an isolated quantum system a description of macroscopic equilibrium properties which agrees with classical thermodynamics.

The previous analysis allows also the determination of the typical values of the reduced density matrix for a component. The starting point is eq. (74) after having identified the subsystem with one component so that $E_m^{sub} = e_m$, and the remaining components with the environment. Therefore the summation at the r.h.s. of eq. (74) leads to the number of states for a system of $(n-1)$ components with a cut-off energy $E_{max} - e_m$:

$$\langle \bar{\mu}_{m,m} \rangle = \frac{N(n-1, E_{max} - e_m)}{N(n, E_{max})} \tag{108}$$

Then, by specifying the number $N$ of states according to the entropy eq. (87) expressed in the thermodynamic limit by means of the entropy per component, $S(n, E_{max}) = ns(E_{max}/n)$ and $S(n-1, E_{max} - e_m) = (n-1)s((E_{max} - e_m)/(n-1))$, we obtain the typical reduced density matrix as a function of the number of components and of the cut-off energy per components eq. (89)

$$n \to \infty: \quad \ln \langle \bar{\mu}_{m,m} \rangle = \frac{n-1}{k_B} s\left( \frac{n e_{\max} - e_m}{n-1} \right) - \frac{n}{k_B} s(e_{\max}) =$$
$$= s'(e_{\max}) \frac{e_{\max} - e_m}{k_B} - \frac{s(e_{\max})}{k_B} \tag{109}$$

where the r.h.s. is the leading term obtained as series expansion about $n \to \infty$. By eliminating the entropy per component and its derivative according to eqs. (96), (97) and (105), we finally obtain

$$n \to \infty: \quad \langle \bar{\mu}_{m,m} \rangle = \frac{\exp(-e_m / k_B T)}{Q} \tag{110}$$

which is precisely the probability eq. (106) for the component states.

## VI - CONCLUDING REMARKS

In this paper we have demonstrated that a complete and self-consistent statistical treatment is possible for isolated quantum systems starting from its time evolution described by the Schrödinger equation, and leading to a description of equilibrium properties in agreement with classical thermodynamics. Such a treatment is based on two fundamental building blocks. The first is the Pure State Distribution (PSD) resulting from the time evolution of a given realization of the isolated quantum system. It requires the confinement of the wavefunction to a finite dimensional space, the so called active Hilbert space defined on the basis of an upper energy cut-off $E_{\max}$, and it allows the identification of the equilibrium distribution of any property on the basis of the time evolution of the quantum state. The same structure of the Schrödinger equation allows the identification of the constants of motion as the populations along the Hamiltonian principal directions. Therefore the equilibrium properties depend on the realization of the isolated quantum system through the set of populations. Population dependent functions are also introduced for the microscopic definitions of the internal energy and of the entropy in correspondence of the expectation value of the Hamiltonian and of the Shannon entropy for the populations, respectively.

Thus, as long as the populations of a given isolated system are unknown, a second building block has to be introduced for an ensemble characterizing the statistical distribution of the possible population sets. Since the information content of the Schrödinger equation has been already employed in the derivation of the Pure State Distribution, quantum mechanics is not helpful in characterizing the statistical ensemble for the populations. It has to be postulated on the basis on some a priori constraints, like symmetries of the population distribution, and subsequently it has to be validated from the predictions with the known behaviour of material systems. To this aim we have employed the Random Pure State Ensemble (RPSE) obtained from an homogeneous distribution of the wavefunction within the unit sphere in the active Hilbert space. Because of its symmetry, RPSE allows a rather simple characterization of the marginal distribution of the populations and of their correlations. On this basis we have evaluated the variance within RPSE of relevant properties like the microscopic internal energy, the microscopic entropy and the equilibrium reduced density matrix of a subsystem, leading to the proof of their typicality for an increasing dimension of the active space. In this limit the population dependence of microscopic equilibrium properties can be neglected, and their macroscopic values can be identified with the corresponding RPSE averages. In this way one identifies the fundamental thermodynamic properties like the internal energy and the entropy.

In order to demonstrate that such an approach is consistent with the behaviour of material systems, one has to recover the classical thermodynamics and, in particular, the fundamental equation of state connecting the entropy and internal energy of system at constant volume. To this purpose one has to take into account that both PSD and RPSE are defined for a given active space determined by a the cut-off energy $E_{max}$. Therefore all the macroscopic properties are intrinsic functions of such a parameter. By eliminating the $E_{max}$ dependence between the entropy $S$ and the internal energy $U$, the fundamental equation of state $S = S(U)$ is naturally recovered. Furthermore, by considering the isolated quantum system as an ensemble of weakly interacting components, it is shown for generic systems that both $S$ and $U$ are extensive properties, and that $S(U)$ is an increasing convex function of $U$. In this way the usual thermal description of material systems is recovered since the temperature $T = dU/dS$ results to be a positive intensive parameter increasing with the internal energy. Moreover it has been rigorously shown that typical reduced density matrix of a component takes the form of the canonical distribution at the temperature determined by the $S(U)$ equation of state.

Then the following question arises: how to collocate the standard microcanonical statistics in such a framework? As a matter of fact in the microcanonical approach from the beginning one attributes to the system a fixed set of populations, more precisely identical populations within an energy shell between the boundaries $E_{\min}$ and $E_{\max}$, with a small enough width $E_{\max} - E_{\min}$. The superiority, shared by other typicality based approaches, of the present treatment derives from the recognition of microscopic parameters, like the populations, which distinguish the different possible realizations of the isolated quantum system. Then a satisfactory analysis requires, instead of attributing to them a priori values, a probabilistic description by means of a suitable statistical ensemble. In this framework the typicality plays a fundamental methodological role in order to establish a self-consistent description of the macroscopic world. By demonstrating that the relevant microscopic properties have a negligible variance with the ensemble, one recovers macroscopic observables identified with the typical values which are independent of the microscopic details of the realization of the quantum system.

Our analysis is base on a specific model, the RPSE, for the statistical ensemble of populations. We recall the fundamental work of Popescu et al. [11] demonstrates that typicality, at least for the reduced density matrix of a subsystem, is a general feature of random distributions of the wavefunction. Basically different ensemble models can be discriminated on the basis of their predictions of macroscopic properties and, in particular, of the agreement with classical thermodynamics. Such a constraint, however, is not sufficient to identify univocally the statistical ensemble. For instance one can introduce an alternative form of RPSE by considering as active Hilbert space the Hamiltonian principal directions with energies $E_k$ within an interval defined by both a lower boundary $E_{\min}$ and an upper boundary $E_{\max}$, $E_{\min} \leq E_k < E_{\max}$, like in the standard microcanonical statistics. The random distribution of quantum states within such an energy shell is the basic assumption of the typicality analysis of a subsystem by Goldstein et al. [12]. Our treatment can be applied also to such a type of ensemble by deriving, however, internal energy $U$ and entropy $S$ depending on both the boundaries $E_{\min}$ and $E_{\max}$. Then the fundamental equation of state $S = S(U)$ is no more derivable by direct elimination of the parameters defining the active space, unless one assumes from the beginning a relation between the two boundaries, for instance that $E_{min}$ is a fixed fraction of $E_{max}$. The requirement of methodological simplicity or, in other words, the "Occam razor" clearly supports our choice of RPSE with the cut-off energy $E_{max}$ only determining the active space. On the other hand we think that the issue of the more appropriate statistical ensemble is still an open field of research, even if the present work shows that RPSE provides a self-consistent treatment of isolated quantum systems in agreement with classical thermodynamics.

**Appendix A: Asymptotic estimates of RPSE averages.**

In this Appendix we will derive the leading contributions, with respect to an infinite dimension $N$ of the active Hilbert space, for RPSE averages of particular functions of scaled populations eq. (59), which are required in the calculation of the entropy variance reported in Section III. Let us first consider the average of a function $g(x_1)$ of the first scaled population

$$\langle g(x_1) \rangle = \int_0^1 dP_1 g(P_1 N) p(P_1) = \int_0^N dx_1 g(x_1) \frac{p(x_1/N)}{N} \tag{A1}$$

The $N$-dependent function in the integral, according to eq. (40) for the probability density of one population, can be written as

$$\frac{p(x_1/N)}{N} = (1-1/N)(1-x_1/N)^{N-2} = (1-1/N)\exp\{(N-2)\ln(1-x_1/N)\} \tag{A2}$$

From the series expansion of the logarithmic function,

$$\ln(1-x_1/N) \simeq -x_1/N - x_1^2/2N^2 \tag{A3}$$

by retaining the contributions up to $1/N$ at the exponent

$$\frac{p(x_1/N)}{N} \simeq e^{-x_1}(1-1/N)\exp\{(2x_1 - x_1^2/2)/N\} \tag{A4}$$

and finally, from the series expansion of the $N$-dependent exponential function, we get the following expansion up to $1/N$

$$\frac{p(x_1/N)}{N} \simeq e^{-x_1}\left\{1 - \frac{1-2x_1 + x_1^2/2}{N}\right\} \tag{A5}$$

which, by substitution into eq. (A1), leads to eq. (61) once the upper integration boundary is brought to infinity.

In a similar way one can derive eq. (66) as the asymptotic estimate of the RPSE average

$$\langle g(x_1)g(x_2)\rangle = \int_0^1 dP_1 \int_0^{1-P_1} dP_2\, g(P_1 N)g(P_2 N) p(P_1,P_2) = \int_0^N dx_1 \int_0^{N-x_1} dx_2\, g(x_1)g(x_2) \frac{p(x_1/N, x_1/N)}{N^2} \quad \text{(A5)(A6)}$$

Like in eq. (A2), by inserting eq. (42) for the probability density of two populations, the $N$-dependent function in the integral can be rewritten as

$$\begin{aligned}\frac{p(x_1/N, x_2/N)}{N^2} &= (1-1/N)(1-N/2)\left(1-\frac{x_1+x_2}{N}\right)^{N-3} = \\ &= (1-1/N)(1-N/2)\exp\{(N-3)\ln[1-(x_1+x_2)/N]\}\end{aligned} \quad (A7)$$

By using the same expansion of eq. (A3) for the logarithmic function, and by retaining the terms up the first order in $1/N$, we get the following approximation

$$\begin{aligned}\frac{p(x_1/N, x_2/N)}{N^2} &\simeq e^{-x_1}e^{-x_2}(1-1/N)(1-2/N)\exp\{3(x_1+x_2)/N - (x_1+x_2)^2/2N\} \simeq \\ &\simeq e^{-x_1}e^{-x_2}\left[1-(3-3x_1-3x_2+x_1 x_2+x_1^2/2+x_2^2/2)/N\right]\end{aligned} \quad (A8)$$

which, after substitution into eq. (A5), leads to eq. (66) by bringing to infinity the upper integral boundaries. By means of explicit values of the following integrals [ref]

$$\int_0^\infty dx\, x\ln x = 1-\gamma \qquad \int_0^\infty dx\, x^2 \ln x = 3-2\gamma \qquad \int_0^\infty dx\,(x\ln x)^2 = 2\gamma^2 - 6\gamma + 2 + \pi^2/3 \quad (A9)$$

where $\gamma$ is the Euler constant, the numerical estimate eq. (68) is recovered.

**Appendix B: Asymptotic estimate of the number $N$ of states.**

First we derive an approximate form of the density of states $G(n,E)$ of eq. (85) when the number $n$ of components is very large albeit finite. First we consider the case of components having a finite dimensional energy spectrum $(e_0, e_1, ..., e_M)$ such that the occupation numbers $i_m$ are restricted to $m = 0 \div M$. Since by increasing the number $n$ of components each of the occupation numbers becomes larger and even larger, one is legitimated to approximate the factorials according to the Stirling formula [38]

$$j! \cong \sqrt{2\pi}\,\frac{(j+1)^{j+1/2}}{e^{j+1}} \tag{B1}$$

Correspondingly, by assigning the real positive axis to the domain for the occupation numbers, the summation on each occupation number can be approximated by an integral, so that for a given function $f(i)$ of the set of occupation numbers one can employ the approximation

$$\sum_{i} f(i)\delta_{n,\sum_{m=0}^{M}i_m} = \int di\, f(i)\delta\left(n-\sum_{m=0}^{M}i_m\right) := \prod_{m=1}^{M}\left(\int_0^\infty di_m\right)f(i)\delta\left(n-\sum_{m=0}^{M}i_m\right) \tag{B2}$$

where the Kronecker symbol has been substituted by the corresponding Dirac delta function. Then, by applying such an approximation to the density of states eq. (85) one obtains the following relation

$$G(n,E) = \int di\,\delta\left(n-\sum_{m=0}^{M}i_m\right)\delta\left(E-\sum_{m=0}^{M}e_m i_m\right)\sqrt{e}\left(\frac{e}{\sqrt{2\pi}}\right)^{M+1/2}\frac{(n+1)^{n+1/2}}{\prod_{m=0}^{M}(i_m+1)^{i_m+1/2}} \tag{B3}$$

In order to facilitate further elaborations, it is convenient to employ integration variables $q_m \equiv i_m/n$ which are normalized with respect to the number $n$ of components, so that the density of states with respect to energy $e \equiv E/n$ per components reads

$$G(n,ne) = \int dq\,\delta\left(1-\sum_{m=0}^{M}q_m\right)\delta\left(e-\sum_{m=0}^{M}e_m q_m\right)e^{D(n,q)} \tag{B4}$$

where $q=(q_0,q_1,\cdots,q_M)$ and we have introduced the function

$$D(n,q) = 1/2 + (M-1)\ln n + (M+1/2)\ln\left(e/\sqrt{2\pi}\right) + (n+1/2)\ln(n+1) - \\ -\sum_{m=0}^{M}(nq_m+1/2)\ln(nq_m+1) \tag{B5}$$

Notice that for $n\to\infty$ and fixed $q$, the function $D(n,q)$ has the following leading contribution proportional to the number $n$ of components

$$n\to\infty:\quad D(n,q) = -n\sum_m q_m \ln q_m \tag{B6}$$

while the other contributions are at most of the order of $\ln n$.

In order to evaluate the integral of eq. (B4), we introduce the quadratic expansion of the function $D(n,q)$ about its maximum to be derived under the two constraints

$$\sum_{m=0}^{M} q_m = 1 \qquad \sum_{m=0}^{M} e_m q_m = e \tag{B7}$$

By employing two Lagrange multipliers $\alpha$ and $\beta$ (to be considered as functions of both the number $n$ of components and of the energy $e$ per component) for the constraints eq. (B6), the associated function

$$F(n,e,q) = D(n,q) - n\alpha(n,e)\sum_{m=0}^{M} q_m - n\beta(n,e)\sum_{m=0}^{M} e_m q_m \tag{B8}$$

allows the calculation of the maximum location $\hat{q}(n,e)$ as solution of the equation

$$\left.\frac{\partial F(n,e,q)}{\partial q_m}\right|_{q=\hat{q}(n,e)} = 0 \quad \forall m \tag{B9}$$

with the Lagrange multipliers determined by imposing to the solution $\hat{q}(n,e)$ the constraints eq. (B8). Explicit solutions of such a problem can be derived in the limit $n \to \infty$ by replacing into eq. (B8) $D(n,q)$ with its asymptotic form $D^{\infty}(n,q)$ of eq. (B6). Then maximum location is found in the form

$$n \to \infty: \quad \hat{q}(n,e) = p(e) \tag{B10}$$

with parameters $p_m(e)$ defined by eq. (92) and independent of the number $n$ of components. Notice that eq. (93) derives from the second of constraints eq. (B8).

Once the maximum location $\hat{q}(n,e)$ has been identified, one can perform the quadratic expansion of $D(n,q)$ about its maximum, which reads as

$$D(n,q) = \hat{D}(n,e) + \frac{1}{2}\sum_{m,m'} \hat{D}^{(2)}_{m,m'}(n,e)[q_m - \hat{q}_m(n,e)][q_{m'} - \hat{q}_{m'}(n,e)] \tag{B11}$$

with

$$\hat{D}(n,e) := D(n,\hat{q}(n,e)) \tag{B12}$$

and

$$\hat{D}^{(2)}_{m,m'}(n,e) := \left.\frac{\partial^2 D(n,q)}{\partial q_m \partial q_{m'}}\right|_{q=\hat{q}(n,e)} = -n\frac{\hat{q}_m(n,e) + 3/2n}{[\hat{q}_m(n,e) + 1/n]^2}\delta_{m,m'} \tag{B13}$$

All the principal curvatures of $D(n,q)$ at the minimum are negative, and they are proportional to the number $n$ of components for $n \to \infty$ when $\hat{q}(n,e) = p(e)$. Therefore one expects for $e^{D(n,q)}$ a bell shaped profile with widths decreasing for $n \to \infty$, and this justifies the substitution of $D(n,q)$ with its quadratic expansion eq. (B11). Correspondingly one can supply the following estimate for the density of states eq. (B4)

$$G(n,ne) = e^{\hat{D}(n,e)} \hat{I}(n,e) \tag{B14}$$

with the integral

$$\hat{I}(n,e) = \int dq\, \delta\left(1 - \sum_{m=0}^{M} q_m\right) \delta\left(e - \sum_{m=0}^{M} e_m q_m\right) \exp\left\{\sum_{m=0}^{M} \hat{D}^{(2)}_{m,m}(n,e)[q_m - \hat{q}_m(n,e)]^2 / 2\right\} \tag{B15}$$

Since the function to be integrated is a multidimensional Gaussian with eq. (B13) determining the widths, one can conclude that in the thermodynamic limit, $n \to \infty$, the $\hat{I}(n,e)$ function has a power law dependence on the number of components.

Given the previously derived density of states, we can now evaluate the dimension of the active Hilbert space from eq. (86)

$$N(n, ne_{\max}) = n \int_0^{e_{\max}} de\, \hat{I}(n,e) e^{\hat{D}(n,e)} \tag{B16}$$

Let us assume as an hypothesis to be verified a posteriori that, for $n \to \infty$, $\hat{D}(n,e)$ is an asymptotically increasing function of the energy parameter $e$. Then we can introduce a linear energy expansion about the upper integral boundary

$$\hat{D}(n,e) = \hat{D}(n,e_{\max}) + \hat{D}'(n,e_{\max})(e - e_{\max}) \tag{B17}$$

where $\hat{D}'(n,e) \equiv \partial \hat{D}(n,e)/\partial e$. Correspondingly, by evaluating the pre-exponential factor $\hat{I}(n,e)$ at the upper integral boundary $e_{\max}$ and by extending to $-\infty$ the lower integral boundary, from eq. (B16) we obtain the following explicit result

$$N(n, ne_{\max}) = n\hat{I}(n,e_{\max}) \frac{e^{\hat{D}(n,e_{\max})}}{\hat{D}'(n,e_{\max})} \tag{B18}$$

Then, by inserting into eq. (87), we get the explicit form of the entropy

$$S(n, ne_{\max}) = k_B \hat{D}(n, e_{\max}) + k_B \ln \frac{n\hat{I}(n, e_{\max})}{\hat{D}'(n, e_{\max})} \qquad (B19)$$

where according to eq. (B12), eq. (B10) and eq. (B6)

$$n \to \infty: \quad \hat{D}(n, e_{\max}) = -n \sum_m p_m(e_{\max}) \ln p_m(e_{\max}) \qquad (B20)$$

The first term at the r.h.s. of eq. (B19) in the thermodynamic limit is proportional to the number of components, while the second term brings at most a logarithmic contribution on the number of components, and therefore it can be neglected. This allows the calculation of the limit eq. (90) for the entropy so recovering eq. (91). On the other hand, such a result justifies the starting hypothesis since, for $n \to \infty$, $\hat{D}(n,e) = ns(e)/k_B$ increases with the energy like the entropy, with a slope proportional to the number of components.

The previous analysis can be generalized to components having an infinite dimensional spectrum. In the case of the truncation of the single component spectrum according to the maximum allowed energy $e_M$, one recovers from the previous treatment eqs. (91)-(93) with the summation on the $m$ index from 0 to $M$. Then the limit $M \to \infty$ can be performed which is equivalent of removing the truncation on the single component energy spectrum. It should be evident, for instance from eq. (92) for $Q(e_{\max})$, that the restriction of the energy eigenvalues of the components to a finite set does not modify substantially the results as long as the condition $\beta(e_{\max})e_M \gg 1$ is satisfy by the largest allowed eigenvalue $e_M$.

**Acknowledgements**

The authors acknowledge the support by Univesità degli Studi di Padova through 60% grants.

**Notes and References**